\documentstyle[12pt,psfig]{article}

\newcommand{\be}{\begin{equation}}
\newcommand{\ee}{\end{equation}}
\newcommand{\ba}{\begin{eqnarray}}
\newcommand{\ea}{\end{eqnarray}}

\begin{document}

\title{Kirkwood Phase Transition for Boson \\
 \vspace{-0.2cm} and Fermion Hard-Sphere Systems}
\author{ M.A. Sol\'{\i}s$^{1,2}$, M. de Llano$^{3}$ and J.W. Clark$^{1}$ \\ 
\vspace{-0.2cm} $^{1}$ Department of Physics, Washington University \\
\vspace{-0.2cm} St. Louis, Missouri 63130, USA \\
\vspace{-0.2cm} $^{2}$ Instituto de F\'{\i}sica, UNAM, Apdo. Postal 20-364, \\
\vspace{-0.2cm} 01000 M\'exico, D.F., Mexico \\
\vspace{-0.2cm} $^{3}$ Instituto de Investigaciones en Materiales, UNAM, \\
Apdo. Postal 70-360, 04510 M\'{e}xico, DF, Mexico \\
}
\maketitle

\begin{abstract}
The London ground-state energy formula as a function of number density $\rho 
$ for a system of boson hard spheres of diameter $c$ at zero temperature
(corrected for the reduced mass of a pair of particles in a
``sphere-of-influence'' picture) generalized to describe fermion hard-sphere
systems with four and two intrinsic degrees of freedom such as $^{3}$He or
neutron matter and symmetric nuclear matter, respectively, is proposed as
the crystalline energy branch for hard-sphere systems. For the fluid branch
we use the well-known, exact, low-density equation-of-state expansions for
many-boson and many-fermion systems, appropriately extrapolated to physical
densities. Here, via a double-tangent construction the crystallization and
melting densities for boson and fermion hard spheres are determined. They
agree well with variational Monte Carlo, density-functional, and Green
Function Monte Carlo calculations. \newline

\end{abstract}


\section{Introduction}

As in the classical hard-sphere system where a ``freezing'' transition from
a fluid to a crystalline phase was found \cite{Kirw}\ to occur as one
increased density, one expects the same phenomenon to occur in quantum
systems, whether of bosons or fermions.

An analytical formula for the ground state energy $E$ of a boson hard-sphere
system for all densities $\rho $ was proposed by London \cite{London} as 
\begin{equation}
E/N=\frac{2\pi \hbar ^{2}c}{m}\frac{1}{(\rho ^{-1/3}-\rho _{0}^{-1/3})^{2}}%
\frac{1}{(\rho ^{-1/3}+b\,\rho _{0}^{-1/3})},  \label{london}
\end{equation}%
where $m$ is the particle mass, $c$ is the hard-sphere diameter, with $\rho
_{0}\equiv \sqrt{2}/c^{3}$ the ultimate density \cite{Rogers} at which a
system of identical classical hard spheres close packs in some
primitive-hexagonal arrangement, e.g., face-centered-cubic, and $%
b=(2^{5/2}/\pi )-1$. The justification given for Eq. (\ref{london}) is that
it reduces smoothly to well-known limiting expressions at both low and high
densities, namely 
\begin{equation}
E/N \; \smash {\ \mathop{\relbar\joinrel\longrightarrow}\limits_{\rho \to
0}\ \ }\;(2\pi \hbar ^{2}/m)\rho \,c,  \label{tocero}
\end{equation}%
\begin{equation}
E/N\;\smash {\ \mathop{\relbar\joinrel\longrightarrow}\limits_{\rho \to
\rho_0}\ \ }\;A\;(\hbar ^{2}/2m)(\rho ^{-1/3}-\rho _{0}^{-1/3})^{-2}.
\label{tocp}
\end{equation}%
Here $A=\pi ^{2}/2^{1/3}\simeq 7.834$ is a constant called the {\it residue }%
of the pole at close packing. The low-density leading term (\ref{tocero}) is
the celebrated Lenz \cite{Lenz} term, calculated by him as the leading
correction to the energy due to an ``excluded volume'' effect. The limiting
case Eq. (\ref{tocp}) comes from the lowest Schr\"{o}dinger equation
eigenvalue of a particle in the spherical cavity.\ It is just the kinetic
energy of a point particle of mass $m$ inside a spherical cavity of radius $%
r-c$, where $r$ is the average separation between two neighboring hard
spheres and $r=(\sqrt{2}/\rho )^{1/3}$ by assuming a primitive-hexagonal
arrangement for the cavities.

It was found \cite{Solis} that the derivation of the high-density extreme of
the original \cite{London} (boson) London equation (\ref{london}) contains 
{\it one fundamental error}: the spherical cavity of radius $r-c$ alluded to
above in reality refers to a ``sphere of influence'' of {\it two} particles.
Thus, the particle mass used in obtaining (\ref{tocp}) {\it should refer to
the reduced mass} $m/2$. This yields the constant 
\begin{equation}
b\equiv 2^{3/2}/\pi -1  \label{bnew}
\end{equation}%
instead of \thinspace\ $2^{5/2}/\pi -1$ \thinspace\ given by London (\ref%
{london}). The result (\ref{london}) with (\ref{bnew}) is designated the 
{\it modified London (ML) equation}, which continues to satisfy (\ref{tocero}%
) as this is independent of the constant $b$. The residue {$A$ in (\ref{tocp}%
) then becomes $2^{2/3}\pi ^{2}\simeq 15.7$ } and fully agrees with the
empirical residue of 15.7 $\pm $ 0.6, extracted by Cole \cite{Cole} from
high-pressure crystalline-branch data in $^{3}$He, $^{4}$He, H$_{2}$ and D$%
_{2}$ systems. Moreover, this ML equation agrees dramatically better than
the original London (L) equation with Green Function Monte Carlo (GFMC) \cite%
{GFMC} computer-simulation datapoints for both fluid and crystalline
branches of the boson hard-sphere system.

A generalized London equation has also been proposed \cite{Ren} for $N$-{\it %
fermion} hard-sphere systems, with $\nu $ intrinsic degrees of freedom for
each fermion. Two differences appear with respect to the boson London
formula: a) unlike the boson case, the ground state kinetic energy is
nonzero and is added as a well-known \cite{FW} $\nu $-dependent leading
term, and b) the constant $b$ is allowed to be $\nu $-dependent, namely $%
b(\nu )=[(\nu -1)/\nu ](b+1)-1.$This form ensure a $\nu $-{\it independent}
energy at close-packing where, since the spheres can be labelled
indistinguishability and thus particle statistics are lost as expected in
this classical limit. Replacing the value of the constant $b$ by the new one
(\ref{bnew}) gives a {\it generalized modified London equation} (ML$_{\nu }$%
) 
\begin{equation}
E/N=C_{\nu }\,\rho ^{2/3}+\bigg(\frac{\nu -1}{\nu }\bigg)\frac{2\pi \hbar
^{2}c}{m}\frac{1}{(\rho ^{-1/3}-\rho _{0}^{-1/3})^{2}}\frac{1}{[\rho
^{-1/3}+b(\nu )\,\rho _{0}^{-1/3}]}  \label{gen}
\end{equation}%
with 
\begin{equation}
C_{\nu }\equiv \frac{3\hbar ^{2}}{10m}\bigg(\frac{6\pi ^{2}}{\nu }\bigg)%
^{2/3}\smash {\ \mathop{\relbar\joinrel\longrightarrow}\limits_{\nu \to
\infty}\ \ }0
\end{equation}%
where in the limit $\nu \rightarrow \infty $ the new constant $b(\nu
)\rightarrow b$ and (\ref{gen}), i.e., goes over into the boson case stated
above. The low-density limit of Eq. (\ref{gen}) becomes 
\begin{equation}
E/N\smash {\ \mathop{\relbar\joinrel\longrightarrow}\limits_{\rho \to 0}\ \ }%
C_{\nu }\,\rho ^{2/3}+\bigg(\frac{\nu -1}{\nu }\bigg)\frac{2\pi \hbar ^{2}}{m%
}\rho c,
\end{equation}%
where the second term on the rhs is the Lenz term for $\nu $-component
fermions. On the other hand, for $\rho \rightarrow \rho _{0}\equiv \sqrt{2}%
/c^{3}$ one sees that (\ref{gen}) reduces to (\ref{tocp}) as it should. In
other words, hard-sphere fermions, bosons or ``boltzons'' {\it must close
pack at the same density}. From this it follows that the residue for bosons
or fermions is the same and equal to $2^{2/3}\,\pi ^{2}\simeq 15.7$, in line
with Ref. \cite{Cole}.

In addition to the Lenz term (\ref{tocero}) for the low-density fluid
branch, several higher-order corrections to the ground state energy per
particle have been derived using quantum field-theoretic many-boson
perturbation theory \cite{boson}, namely 
\begin{equation}
E/N=\frac{2\pi \hbar ^{2}\rho c}{m}\bigg\{1+C_{1}(\rho
c^{3})^{1/2}+C_{2}\rho c^{3}\,\ln (\rho c^{3})+C_{3}\rho c^{3}+o(\rho c^{3})%
\bigg\},  \label{bos}
\end{equation}%
where $C_{1}=128/15\sqrt{\pi }$, $C_{2}=8(4\pi /{3}-\sqrt{3})$, but $C_{3}$
is an as yet unknown constant. Here $c$ is the S-wave scattering length in
general, and becomes the sphere diameter for a hard-core potential. The
series is clearly not a power series expansion, and is at best an asymptotic
series.

Similarly, for an $N$-fermion hard-sphere system the corresponding series is %
\cite{fermion} 
\begin{eqnarray}
E/N &=&\frac{3}{5}\frac{\hbar ^{2}k_{F}^{2}}{2m}\bigg\{%
1+C_{1}(k_{F}c)+C_{2}(k_{F}c)^{2}  \nonumber \\
&+&[C_{3}r_{0}/2c+C_{4}A_{1}(0)/c^{3}+C_{5}](k_{F}c)^{3}+C_{6}(k_{F}c)^{4}%
\ln (k_{F}c)  \nonumber \\
&+&[C_{7}r_{0}/2c+C_{8}A_{0}^{\prime \prime
}(0)/c^{3}+C_{9}](k_{F}c)^{4}+o(k_{F}c)\bigg\},  \label{fer}
\end{eqnarray}%
where the $C_{j}$ ($j=1,2,...,9$) are dimensionless coefficients depending
on $\nu $ and are given in Ref. \cite{coeff} for $\nu =2$ and $\nu =4$. The
Fermi momentum $\hbar k_{F}$ is defined through the particle number density $%
\rho \equiv N/\Omega =\nu k_{F}^{3}/6\pi ^{2}$, with $\Omega $ the system
volume.

Unfortunately, both low-density expansions (\ref{bos}) and (\ref{fer}) lack
accuracy at moderate to high densities, including the saturation (or
equilibrium, zero-pressure) densities of liquid $^{4}$He ($\nu =\infty $)
and liquid $^{3}$He ($\nu =2$) or nuclear matter ($\nu =4$). However, one
can extrapolate the series for hard-sphere systems to physical and even to
close-packing densities through the use of Pad\'{e} and so-called
``tailing'' \cite{tailing} approximants. This method, called Quantum
Thermodynamic (or Van der Waals) Perturbation Theory (QTPT) \cite{QTPT}, has
provided fairly accurate representations of the {\it fluid} branch of the
equation of state \cite{Annals} beyond the presumable phase transition
densities but without enough credibility as one approaches close packing.
This is clear since one does not possess a ground-state energy function with
implicit information of {\it both} fluid and crystalline branches. Thus, for
the {\it crystalline} branch we employ the generalized modified London
equations (\ref{london}) which reproduce the correct closest packing density
value $\rho =\sqrt{2}/c^{3}$ as well as good behavior near closest packing
as suggested by a correct residue value. For the {\it fluid} branch we use
the Pad\'{e} extrapolations based on the series (\ref{bos}) and (\ref{fer}).

This paper unfolds as follows. In Sections 2 and 3 we construct the fluid
branches for hard-sphere bosons and fermions, respectively. In Sec. 4 we use
them together with their respective modified London equation crystalline
branches to determine, via a double-tangent construction ensuring equality
of pressure in both phases, the melting and crystallization densities as
well as energy and density changes. In Sec. 5 we state our conclusions.

\section{Boson hard-sphere fluid branch}

In order to extrapolate to higher densities the low-density series (\ref{bos}%
), we define 
\begin{equation}
E/N=\frac{2\pi \hbar ^{2}}{m}\rho ce_{0}(x),\quad x\equiv (\rho c^{3})^{1/2},
\end{equation}%
with 
\begin{equation}
e_{0}(x)\equiv 1+C_{1}x+C_{2}x^{2}\,\ln x^{2}+C_{3}x^{2}+O(x^{3}\,\ln x^{2}).
\label{e0b.a}
\end{equation}%
Alternatively, one can analyze the series 
\begin{equation}
e_{0}^{-1/2}(x)=1+F_{1}\,x+F_{2}\,x^{2}\,\ln x^{2}+F_{3}\,x^{2}+O(x^{3}\,\ln
x^{2})  \label{e0b.b}
\end{equation}%
for $x\ll 1$, where the $F$'s are expressible in terms of the $C$'s, with $%
C_{3}$ and consequently $F_{3}$ unknown. Values of the $C$'s and $F$'s are
given in Table 1. We analyze $e_{0}^{-1/2}(x)$ instead of $e_{0}(x)$ to
ensure that zeros in its extrapolants $\epsilon _{0}^{-1/2}(x)$ are
second-order poles in energy as we expect at close packing (CP). Then we
examine the twelve extrapolants \cite{formas} from the series (\ref{e0b.b})
with three terms. Fitting extrapolants to go through the four GFMC data
points \cite{GFMC} allows estimating a good value for the last coefficient $%
F_{3}$. Forms VII and XI both satisfied these conditions but the mean square
deviation with respect to the four GFMC data points was least with form XI.
Thus we use it as our best extrapolant. So the ground state energy for boson
hard spheres is represented (symbol $\doteq $)\ by%
\begin{equation}
E/N\doteq \frac{2\pi \hbar ^{2}}{m}\rho c\epsilon _{0}(x),
\end{equation}%
with 
\begin{equation}
\epsilon _{0}^{-1/2}=\mbox{XI}(x)=\frac{1+F_{2}\,x^{2}\,\ln x^{2}}{%
1-F_{1}\,x-(F_{3}-F_{1}^{2})\,x^{2}}  \label{XIeps0}
\end{equation}%
and $F_{3}=-27.956$. In Fig. 1 this is plotted as a full curve and labeled
XI for the fluid branch. Open circles and squares are GFMC data for fluid
and crystalline branches, respectively. The dashed curve is the Modified
London formula (\ref{london}) representing the crystalline branch. Dots are
Diffusion Monte Carlo (DMC) \cite{Boronat} calculations spanning a wider
range of densities in the fluid region than the GFMC data. From DMC data we
see that although our expression for the fluid branch could be better for
intermediate densities, it agrees well with DMC and GFMC calculations around
the freezing transition. This $\epsilon _{0}^{-1/2}$ predicts a random close
packing (RCP) density $\rho /\rho _{0}=0.776$ which is only ten percent
below classical RCP $\rho /\rho _{0}\simeq 0.86$ \cite{Rogers} expected to be
the ultimate CP density also for quantum hard sphere fluids just from the
fact that particles at CP are localized so that indistinguishability does
not hold and thus statistics are irrelevant.

\begin{figure}[h]
\centerline{\psfig{file=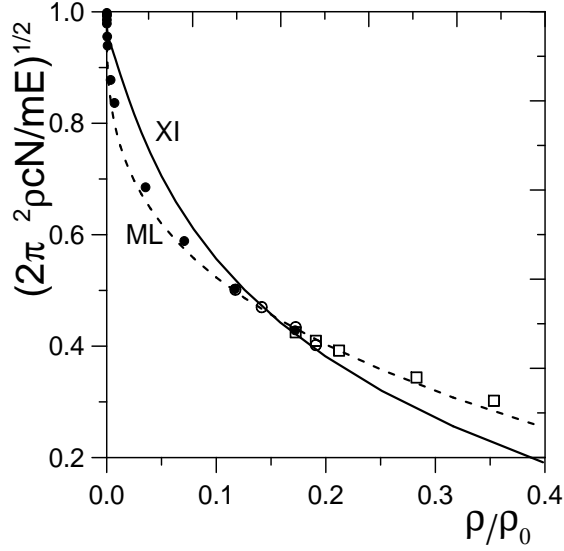,height=3.0in,width=3.0in}} \vspace{%
-0.30cm}
\caption{The quantity $\protect\varepsilon _{0}^{-1/2}=\protect\sqrt{2%
\protect\pi \hbar ^{2}\protect\rho \,c\,N/m\,E}=[1-(\protect\rho /\protect%
\rho _{0})^{1/3}]\protect\sqrt{1+b\,(\protect\rho /\protect\rho _{0})^{1/3}}$
as a function of $\protect\rho /\protect\rho _{0}$ for boson hard sphere
systems: XI is the fluid branch (\ref{XIeps0}), ML is the Modified London
formula (\ref{london}), open circles and squares are GFMC data for the fluid
and crystalline branches respectively, and dots are DMC calculations.}
\end{figure}

\begin{table}[b]
\begin{quotation}
\noindent {\bf Table 1.} For bosons, $C_i$ and $F_i$ coefficients appearing
in Eqs. (\ref{e0b.a}) and (\ref{e0b.b}), respectively. Numbers in quotation
marks are adjusted as indicated in text.
\end{quotation}
\par
\begin{center}
\begin{tabular}{||c|c|c|c||}
\hline\hline
Bosons ($\nu = \infty$) & $i=$1 & 2 & 3 \\ \hline
$C_i$ & 4.81441778 & 19.65391518 & ``73.296" \\ \hline
$F_i$ & -2.40720889 & -9.826957589 & ``-27.956" \\ \hline\hline
\end{tabular}
\\[0pt]
\end{center}
\end{table}

\section{Fermion hard-sphere fluid branch}

For fermion hard-sphere systems we write (\ref{fer}) as 
\begin{equation}
E/N\;=\;\frac{3}{5}\frac{\hbar ^{2}k_{F}^{2}}{2m}\,e_{0}(x);\qquad x\equiv
k_{F}c,\qquad \rho \equiv N/\Omega =\nu k_{F}^{3}/6\pi ^{2},
\end{equation}%
with 
\begin{eqnarray}
e_{0}(x)\,=1 &+&C_{1}\,x+C_{2}\,x^{2}+(C_{3}/3+C_{4}/3+C_{5})\,x^{3} 
\nonumber \\
&+&C_{6}\,x^{4}\,\ln x+(C_{7}/3-C_{8}/3+C_{9})\,x^{4}+o(x^{4}).  \label{fer2}
\end{eqnarray}%
\noindent {\underline{For $\nu =2$}}, $C_{6}=0$ \cite{fermion} so that (\ref%
{fer2}) becomes 
\begin{equation}
e_{0}(x)\,\simeq \,1+D_{1}\,x+D_{2}\,x^{2}+D_{3}\,x^{3}+D_{4}\,x^{4}+o(x^{4})
\label{fer3}
\end{equation}%
for $x\ll 1$. With $D$'s in terms of $C$'s and $\rho =k_{F}^{3}/3\pi ^{2}$.
As in the boson case, instead of $e_{0}$ we write 
\begin{equation}
e_{0}^{-1/2}\simeq
1+F_{1}\,x+F_{2}\,x^{2}+F_{3}\,x^{3}+F_{4}\,x^{4}+F_{5}\,x^{5}+o(x^{5}),
\label{eps0nu2}
\end{equation}%
where the $F_{i}$'s depend algebraically on the $D_{i}$'s in a simple
manner, $F_{5}$ being unknown. Values of $D_{i}$ and $F_{i}$ are given in
Table 2. This series is a simple power series and we use it to construct the
usual Pad\'{e} extrapolants. The approximants to (\ref{eps0nu2}) with four
terms beside the trivial one were analyzed in Ref. \cite{Ho} concluding that
the best was the Pad\'{e} $[4/0](x)$. However, this function does not have a
zero in the region of physical interest, i.e. $0\leq \rho /\rho _{0}\leq 1$
or $0\leq x\leq 3.47$, and so the energy does not manifest a CP as it
should. It was necessary to introduce the fifth term in (\ref{eps0nu2}). Its
five {\it two}-point Pad\'{e} approximants $[3//2](x)$\ were analyzed and $%
F_{5}$ adjusted to ensure a zero. The position of the zero and the
approximant were chosen in such way that the QTPT applied to $^{3}$He with
the Aziz inter-atomic potential reproduces the corresponding GFMC \cite%
{Panoff} data. Eventually, the best extrapolant was the two-point Pad\'{e}
approximant \cite{puertorico} 
\begin{equation}
\epsilon _{0}^{-1/2}\;\doteq \;[3//2](x).
\end{equation}%
Hence $E/N$ becomes 
\begin{equation}
E/N\;=\;\frac{3}{5}\frac{\hbar ^{2}k_{F}^{2}}{2m}\,([3//2](x))^{-2},
\label{efer2}
\end{equation}%
with a CP density $\rho /\rho _{0}=0.732$. The coefficient $F_{5}$ is listed
Table 2 between quotation marks. In Fig. 2 $\varepsilon _{0}^{-1/2}=(3\hbar
^{2}(6\pi ^{2}\rho /\nu )^{2/3}N/10mE)^{1/2}=1+({20\pi (\nu -1)}/{3\nu })({%
2^{1/4}\nu }/{6\pi ^{2}})^{2/3}\{[(\rho /\rho _{0})^{-1/3}-1]^{2}[(\rho
/\rho _{0})^{-1/3}-b(\nu )](\rho /\rho _{0})^{2/3}\}^{-1}$ as a function of $%
\rho /\rho _{0}$ is plotted for fermion hard sphere. For $\nu =2$ the fluid
branch [3//2] (full curve) is close to the numerically exact Ladder (Ladder) %
\cite{Ex.Ladd} (open squares), Variational Fermion-Hypernetted-Chain (VFHNC) %
\cite{VFHNC} (plus sign marks) and L-expansion data \cite{L-expansion} (open
triangles). Fig. 3 is a enlargement of Fig. 2 where we show this excellent
agreement over the whole range of available data. 
\begin{figure}[tbh]
\centerline{\psfig{file=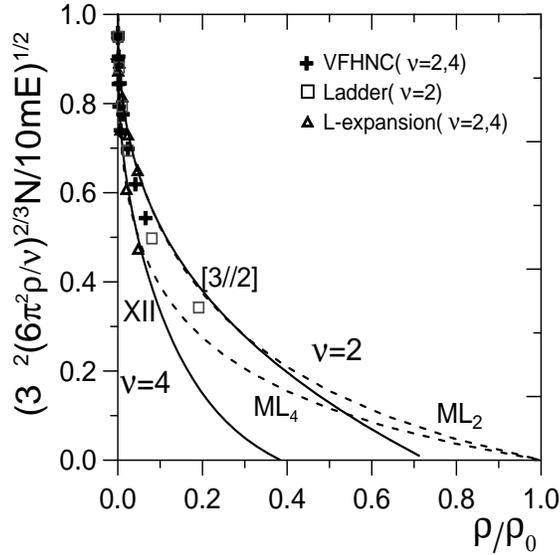,height=3.0in,width=3.0in}}
\vspace{-0.30cm}
\caption{The quantity $\protect\varepsilon _{0}^{-1/2}=(3\hbar ^{2}(6\protect%
\pi ^{2}\protect\rho /\protect\nu )^{2/3}N/10mE)^{1/2}=1+{20\protect\pi (%
\protect\nu -1)}/{3\protect\nu }\times ({2^{1/4}\protect\nu }/{6\protect\pi %
^{2}})^{2/3}\{[(\protect\rho /\protect\rho _{0})^{-1/3}-1]^{2}[(\protect\rho %
/\protect\rho _{0})^{-1/3}-b(\protect\nu )](\protect\rho /\protect\rho %
_{0})^{2/3}\}^{-1}$ as a function of $\protect\rho /\protect\rho _{0}$ for
fermion hard sphere with $\protect\nu =4$ (XII) and $\protect\nu =2$
([3//2]), are full lines. Dashed lines are the corresponding Modified London
formulas.}
\end{figure}

\begin{table}[tbp]
\begin{quotation}
\noindent {\bf Table 2.} For $\nu =2$, $D_i$ and $F_i$ coefficients
appearing in Eqs. (\ref{fer3}) and (\ref{eps0nu2}), respectively. Numbers in
quotation marks are adjusted as is indicated in text.
\end{quotation}
\par
\begin{center}
\begin{tabular}{||c|c|c|c|c|c||}
\hline\hline
$\nu=2$ & $i=$1 & 2 & 3 & 4 & 5 \\ \hline
$D_i$ & 0.353678 & 0.185537 & 0.384145 & -0.024700 & ``-0.265544" \\ \hline
$F_i$ & -0.176833 & -0.045863 & -0.156677 & 0.109672 & ``0.130830" \\ 
\hline\hline
\end{tabular}
\\[0pt]
\end{center}
\end{table}

\noindent {\underline{For $\nu =4$}, Eq. (\ref{fer2}) becomes 
\begin{equation}
e_{0}(x)\,=\,1+D_{1}\,x+D_{2}\,x^{2}+D_{3}\,x^{3}+D_{4}\,x^{4}\,\ln
x+D_{5}\,x^{4}+o(x^{5})  \label{fer4}
\end{equation}%
for $x\ll 1$. }By analogy with bosons and fermions ($\nu =2$), we analyze 
\begin{equation}
e_{0}^{-1/2}=1+F_{1}\,x+F_{2}\,x^{2}+F_{3}\,x^{3}+F_{4}\,x^{4}\,\ln
x+F_{5}\,x^{4}+o(x^{5})  \label{eps0nu4}
\end{equation}%
with all stated $F_{i}$ known. Values of $D_{i}$ and $F_{i}$ are given in
Table 3. Unlike the $\nu =2$ case, this series is not a pure power series as
it contains logarithmic terms. Its tailing approximants are giving in Table
III of Ref. \cite{tailing}. We choose form XII to avoid approximants with
spurious unphysical poles within the interval of physical densities, as well
as to avoid residues falling outside the {\it rigorous} interval 
\begin{equation}
1.63\leq A\leq 27.  \label{Hubbard}
\end{equation}%
This interval was obtained \cite{Hubbard} for regular CP
(face-centered-cubic or hexagonal-close-packing) by generalizing the exact
calculation for a simple cubic lattice based on three mutually perpendicular
linear lattice which gives $A=\pi ^{2}$. 
\begin{figure}[tbh]
\centerline{\psfig{file=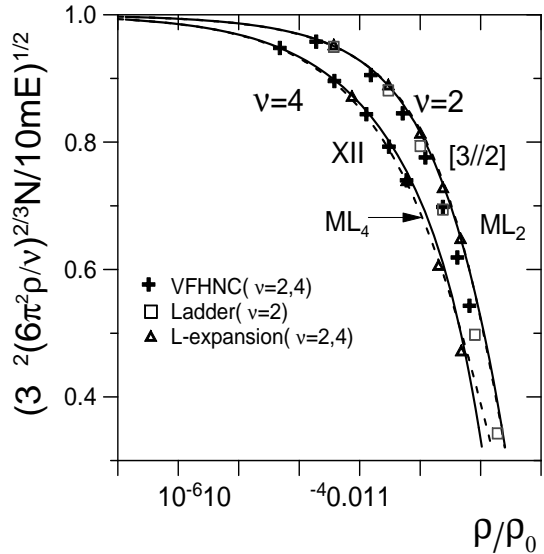,height=3.0in,width=3.0in}} 
\vspace{-0.3cm}
\caption{Enlargement of Fig. 2 at low densities.}
\end{figure}

\begin{table}[tbp]
\begin{quotation}
\noindent {\bf Table 3.} For $\nu =4$, $D_i$ and $F_i$ coefficients
appearing in Eqs. (\ref{fer4}) and (\ref{eps0nu4}), respectively.
\end{quotation}
\par
\begin{center}
\begin{tabular}{||c|c|c|c|c||}
\hline\hline
$\nu=4$ & $i=$1 & 2 & 3 & 4 \\ \hline
$D_i$ & 1.061033 & 0.556610 & 1.300620 & -1.408598 \\ \hline
$F_i$ & -0.530517 & 0.143867 & -0.5806558 & -0.704299 \\ \hline\hline
\end{tabular}
\\[0pt]
\end{center}
\end{table}
Hence, $E/N$ can be written 
\begin{equation}
E/N=\frac{3\hbar ^{2}k_{F}^{2}}{10m}\epsilon _{0}(x),
\end{equation}%
with 
\begin{equation}
\epsilon _{0}^{-1/2}(x)\doteq \mbox{XII}(x)=\frac{%
1+(F_{1}-F_{3}/F_{2})x+(F_{2}-F_{1}F_{3}/F_{2})x^{2}}{%
1-(F_{3}/F_{2})x-F_{4}x^{4}\ln x}.
\end{equation}%
This is plotted in Fig. 2 and 3 as a full curve and labeled XII. We also
plot the corresponding VFHNC data (plus signs) and L-expansion data (open
triangles). In terms of energy, ours are slightly below the VFHNC ones. The
agreement gets better for lower densities. However, the XII approximant is
just above the L-expansion data over the range of densities where they are
available.

\section{Kirkwood phase transition}

Fluid-to-crystal (or freezing) phase transitions for boson hard-sphere
systems have been determined with VMC, GFMC or Density Functional Theory
(DFT) \cite{MWDA}. However, for fermion hard-sphere systems such
calculations have not been done. The purpose of this paper is to provide
very simple but hopefully accurate analytical expressions for the
ground-state energy of quantum hard-sphere systems in general so as to
reproduce crystallization transitions or to give the values for the
parameter involved, as well as other thermodynamic properties such as
pressure and compressibility.

Here, via a double-tangent construction we locate the crystallization $\rho
_{l}$ and melting $\rho _{s}$ number densities for bosons and for fermions
(with $\nu =2$ and$\ 4$). This process is schematically sketched in Fig. 4
for fermions with $\nu =2$, where both fluid and crystalline energies per
particle, in $\hbar ^{2}/mc^{2}$ units, as a function of $\rho /\rho _{0}$,
are shown close to the phase transition. Dots denote crystalline and fluid
phase Maxwell-construction endpoints on the pressure-volume diagram,
obtained by constructing a common tangent to both energy branches ensuring
equality of pressure in both phases.

For boson hard-spheres, in Table 4 we list our calculation (ML) together
with the VMC, GFMC and DFT ones. All of them agree reasonably well with each
other. Here $\Delta (\rho c^{3})$ and $\Delta (E/N)$ signify the change of
density and energy, respectively, between both phases. 
\begin{table}[tbp]
\begin{quotation}
\noindent {\bf Table 4.} Crystallization and melting densities for boson
hard sphere system. NA means not available.
\end{quotation}
\par
\begin{center}
\begin{tabular}{|l|l|l|l|l|}
\hline\hline
Bosons & $\rho_l \, c^3$ & $\rho_s \, c^3$ & $\Delta (\rho_s \, c^3)$ & $%
\Delta (E/N)$ \\ \hline
VMC \cite{VMC} & 0.23 $\pm$ 0.02 & 0.25 $\pm$ 0.02 & 0.02 & 1.28 \\ \hline
DFT \cite{MWDA} & 0.246 & 0.284 & 0.38 & 2.668 \\ \hline
GFMC \cite{GFMC} & 0.25 $\pm$ 0.01 & 0.27 $\pm$ 0.01 & 0.02 & NA \\ \hline
ML & 0.21 $\pm$ 0.02 & 0.23 $\pm$ 0.02 & 0.02 & 1.37 \\ \hline\hline
\end{tabular}
\\[0pt]
\end{center}
\end{table}
For fermion hard spheres with $\nu =4$, e.g., symmetric nuclear matter, we
find a crystallization density given by $\rho _{l}\,c^{3}=0.056$ and a
melting density given by $\rho _{s}\,c^{3}=0.069$. This crystallization
density is of the order of that of matter nuclear (0.32 $\pm $ 0.02 fm$^{-3}$%
) \cite{Anderson,Palmer,Clark1} when we use the hard-sphere radius $c=0.64$
fm determined through the equation of state of Jaqaman {\it et al}. \cite%
{Jaqaman} for nuclear matter interacting through a Skyrme effective
interaction and adjusted to reproduce energy-per-nucleon and saturation
density, i.e., -16 Mev and 0.17 fm$^{-3}$, respectively. 
\begin{table}[tbp]
\begin{quotation}
\noindent {\bf Table 5.} Crystallization and melting densities for fermion ($%
\nu=4, \ 2$) hard sphere system.
\end{quotation}
\par
\begin{center}
\begin{tabular}{|c|c|c|c|c|}
\hline\hline
Fermions & $\rho_l \, c^3$ & $\rho_s \, c^3$ & $\Delta (\rho \, c^3)$ & $%
\Delta (E/N)$ \\ \hline
$\nu=4$ & 0.056 $\pm$ 0.005 & 0.069 $\pm$ 0.006 & 0.013 & 0.29 \\ 
\hline\hline
$\nu=2$ & 0.408 $\pm$ 0.001 & 0.422 $\pm$ 0.001 & 0.014 & 1.589 \\ 
\hline\hline
\end{tabular}
\\[0pt]
\end{center}
\end{table}

For fermion hard spheres with $\nu =2$, e.g., helium three or neutron
matter, we find $\rho _{l}\,c^{3}=0.408$\ at crystallization and $\rho
_{s}\,c^{3}=0.422$\ at melting. These density values agree well with the
experimental ($\rho _{l}\,c^{3}=0.39$, $\rho _{s}\,c^{3}=0.41$) and
calculated \cite{Pand1,Schiff} values for helium three. On the other hand,
taking for the hard-core diameter a value of $c=0.4$ fm, and assuming that
an additional attractive potential would not change the crystallization
density of real neutron matter too much, one could predict a crystallization
density for neutron matter of about 6.38 fm$^{-3}$ (or mass density $m\rho
=10.85\times 10^{15}$ g-cm$^{-3},$ with $m$ the neutron mass) which not only
agrees with a prediction \cite{Canuto} that pure neutron matter crystallizes
when its mass density exceeds 1.5 $\times $ 10$^{15}$ g-cm$^{-3}$ but also
justifies Pandharipande's claim \cite{Clark2} p. 695, of no observation 
of a crystal phase below 4.2 fm$^{-3}$. In Table 5 we summarize our freezing
transition parameters for fermion hard spheres.
\begin{figure}[t]
\centerline{\psfig{file=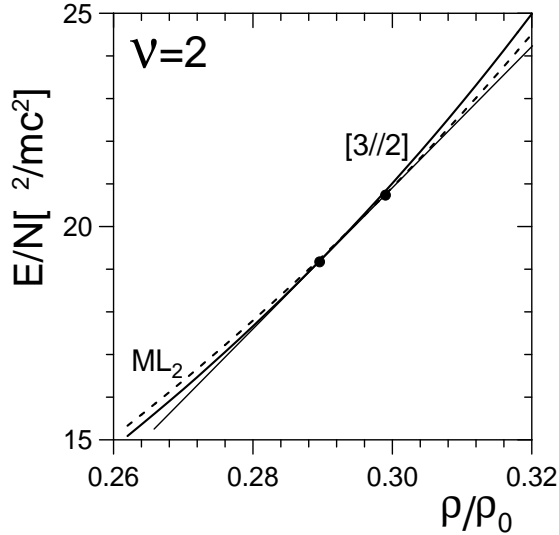,height=3.0in,width=3.0in}} \vspace{%
-0.30cm}
\caption{Energy E/N (in units of $\hbar ^{2}/mc^{2}$) as a function of $%
\protect\rho /\protect\rho _{0}$ for fermion ($\protect\nu =2)$ hard sphere
system and Maxwell double tangent construction. Dots mean crystalline and
fluid edges.}
\end{figure}

\section{Conclusions}

In summary, we have proposed analytical expressions for crystalline and
fluid branches of both boson and fermion quantum hard-sphere systems. We
have been able to calculate the fluid-crystal transition parameters for
bosons and fermions such as symmetrical nuclear matter and neutron matter.
The results support the controversial \cite{Clark2} possibility of
crystallization in nuclear matter as well as in neutron matter. \newline

\noindent{\bf ACKNOWLEDGMENTS}

MAS thanks Washington University for hospitality during a sabbatical year. We also
acknowledge UNAM-DGAPA-PAPIIT (Mexico), grant \# IN106401, and CONACyT
(Mexico), grant \# 27828 E, for partial support.

\pagebreak

\end{document}